# Generation of ring-shaped optical vortices in dissipative media by inhomogeneous effective diffusion

Shiquan Lai[1], Huishan Li[1], Yunli Qui[1], Xing Zhu[2], Dumitru Mihalache[3],

Boris A. Malomed[4], and Yingji He[1,*]

[1]*School of Photoelectric Engineering, Guangdong Polytechnic Normal University,*

*Guangzhou 510665, China*

[2] *Department of Physics and Information Engineering, Guangdong University of*

*Education, Guangzhou 510303, China*

[3] *Horia Hulubei National Institute for Physics and Nuclear Engineering, P.O. Box*

*MG-6, RO-077125, Bucharest-Magurele, Romania*

[4] *Faculty of Engineering, Department of Physical Electronics, School of Electrical*

*Engineering, Tel Aviv University, Tel Aviv 69978, Israel*

*Corresponding author: heyingji8@126.com

**Abstract** By means of systematic simulations we demonstrate generation of a variety of ring-shaped optical vortices (OVs) from a two-dimensional input with embedded vorticity, in a dissipative medium modeled by the cubic-quintic complex Ginzburg-Landau equation with an inhomogeneous effective diffusion (spatial-filtering) term, which is anisotropic in the transverse plane and periodically modulated in the longitudinal direction. We show the generation of stable square- and gear-shaped OVs, as well as tilted oval-shaped vortex rings, and string-shaped bound states built of a central fundamental soliton and two vortex satellites, or of three fundamental solitons. Their shape can be adjusted by tuning the strength and modulation period of the inhomogeneous diffusion. Stability domains of the generated OVs are identified by varying the vorticity of the input and parameters of the inhomogeneous diffusion. The results suggest a method to generate new types of ring-shaped OVs with applications to the work with structured light.

**Keywords** Inhomogeneous effective diffusion, Ginzburg-Landau equation, Optical vortices

## 1 Introduction

In the course of the past three decades, a great deal of interest was drawn to theoretical and experimental studies of the existence, stability, and excitation of localized patterns in optical and matter-wave media [1-13]. This work has produced many findings for temporal, spatial, and spatiotemporal solitons in both conservative and dissipative media, which offer applications to all-optical switching, pattern recognition, parallel data processing, and for guiding light by light. In particular, the cubic-quintic complex Ginzburg-Landau (CGL) equation [14,15] is a generic nonlinear model that predicts the generation of a plethora of temporal, spatial, and spatiotemporal patterns due to the simultaneous balance of gain and loss, and of self-focusing nonlinearity and either diffraction or dispersion (or both of them). The cubic-quintic CGL equation is an adequate model in diverse fields, such as superconductivity and superfluidity, fluid dynamics, reaction-diffusion phenomena, nonlinear photonics, matter waves (Bose–Einstein condensates), quantum field theories, and biophysics. In particular, it predicts diverse species of localized modes, such as fundamental (zero-vorticity) solitons, vortex solitons, soliton clusters, etc., in laser cavities and similar settings [16-41].

Optical vortices (OVs) represent a generic class of self-organized patterns, that have been observed in Kerr-type nonlinear defocusing media [42,43] and display characteristic spiral field profiles with zero field amplitude at the vortex' pivot [44]. However, the robust propagation of OVs is limited by the necessity of the presence of a stable finite-amplitude background field holding the vortices [45-50]. It is worth mentioning applications of OVs to the transfer of optical angular momentum from light to matter, guiding of light by light [51,52], and manipulation and tweezing of nanoparticles in colloidal suspensions [53].

Earlier works have addresses the propagation and interactions of OVs, and of sets of such optical vortices, in various physical settings [54-58]. Recently, results have been reported for solitary vortices supported by localized parametric gain [59,29], localized dark solitons and vortices in defocusing nonlinear media with spatially inhomogeneous nonlinearities [60], guiding and confining of light induced by OVs in cubic-quintic media [61], twin-vortex solitons in nonlocal media [62], and the vortex solitons embedded in flattop Bessel beams [63]. It is also relevant to refer to recent works addressing ring- and elliptic-shaped vortex solitons in relevant physical configurations. The compression and stretching of ring vortices in bulk nonlinear media have been recently investigated both analytically and numerically [64]. The existence and stability of vortex solitons in a ring-shaped partially parity-time-symmetric potential have been studied in detail, revealing that robust nonlinear vortices can be easily created by input Gaussian beams with embedded vorticities [65]. The evolution of elliptic vortex rings in initially quiescent fluids, or under the action of a linear shear flow, was numerically simulated using a lattice Boltzmann method [66].

An essential ingredient of CGL-based models is the term accounting for effective diffusion (spatial filtering, in terms of optics) [67,68]. In particular, it represents ionization of the medium under the action of an intense optical field [69], or relatively poor finesse of the laser cavity [70-72]. Further, the same term is important in models of exciton-polariton condensates, where it represents diffusion of excitons [73-77]. The presence of the spatial filtering is crucially important for the stability of dissipative solitons with embedded vorticity, only zero-vorticity modes being stable in the absence of this term [18,25,33,59,67]. In particular, the creation of robust vortex clusters embedded in two-dimensional beams in media described by the generic cubic-quintic CGL equation with an inhomogeneous effective diffusion term was recently studied [78].

In this work, we numerically investigate generation scenarios for a variety of ring-shaped OVs from a two-dimensional ring-shaped input beam with an embedded vorticity, in the framework of the generic cubic-quintic CGL equation with an inhomogeneous and, generally, anisotropic effective diffusion. In this context, we consider the general case of unequal diffusion coefficients in the two transverse directions, $x$ and $y$, and of unequal periods of modulation in the longitudinal direction, *z*. We demonstrate that the anisotropy and periodic modulation of the effective diffusion makes it possible to create new species of stable dissipative solitons with embedded vorticity, such as square- and gear-shaped ones, oval tilted vortex rings, and strings built of a centrally placed fundamental soliton and two vortex satellites.

The model under consideration is introduced in Section 2. In Section 3 we report generation scenarios for different types of ring-shaped OVs from the input beam with embedded vorticity. Section 4 concludes the work.

## 2 The model

We introduce the cubic-quintic CGL equation of the general form in terms of the spatial-domain optical propagation [78]:

$$iu_z + (1/2)\Delta u + |u|^2 u + \nu |u|^4 u = iR[u], \qquad (1)$$

where $\Delta = \partial^2/\partial x^2 + \partial^2/\partial y^2$ is the transverse Laplacian (*x* and *y* are the transverse coordinates), *z* is the propagation distance, the coefficient in front of the cubic self-focusing term is scaled to be 1, *v* is the quintic self-defocusing coefficient, and $iR[u]$ is a combination of loss and gain terms specified below. In addition to this, Eq. (1) admits a different realization in terms of the spatiotemporal propagation in a planar optical waveguide, with transverse coordinate *x*, while the coordinate *y* is actually the scaled temporal variable, $y \equiv t - z/V_{gr}$, where $V_{gr}$ is the group velocity of the carrier wave, and it is assumed that the group-velocity dispersion has the anomalous sign [3].

The gain and loss terms on the right-hand side of Eq. (1) are represented by $R[u] = \delta u + \varepsilon |u|^2 u + \mu |u|^4 u + \hat{D} u$, where $(-\delta)$ and $(-\mu)$ are the linear and quintic loss coefficients and $\varepsilon$ is the cubic-gain coefficient. Further, operator $\hat{D}$ accounts for the inhomogeneous effective diffusion (spectral filtering), which is periodically modulated in the propagation direction, with different oscillation periods, $T_x$ and $T_y$ and relative amplitudes, $A_x > 0$ and $A_y > 0$, with respect to the corresponding transverse directions $x$ and $y$, respectively:

$$\hat{D} u = \beta \left[ A_x \left| \sin\left(z/T_x\right) \right| \frac{\partial^2 u}{\partial x^2} + A_y \left| \sin\left(z/T_y\right) \right| \frac{\partial^2 u}{\partial y^2} \right] \qquad (2)$$

where $\beta > 0$ is the overall diffusion coefficient. In fact, the spectral filtering of this type can be easier implemented in terms of the above-mentioned spatiotemporal-propagation model, in which the temporal filtering is introduced independently from the spatial diffusion (in the bulk spatial domain, it would be difficult to realize anisotropic spatial diffusion). The periodic modulation of the filtering strength can be introduced as it is done in the broad class of the optical *management* settings [79].

It is also to relevant to mention that, while the particular modulation profile adopted in Eq. (2) is non-smooth, results were also obtained for its smooth counterpart, with the terms in the square brackets replaced by

$A_x \sin^2(z/T_x) \frac{\partial^2 u}{\partial x^2} + A_y \sin^2(z/T_y) \frac{\partial^2 u}{\partial y^2}$ . This modification of the CGL equation produces nearly identical results, up to rescaling of coefficients $A_{x,y}$.

# 3 The generation of a variety of stable ring-shaped optical vortices

The simulations performed for many values of the parameters demonstrate that the generic findings may be adequately represented by results displayed here for $\delta = -0.4$, $\nu = 0.1$, $\mu = 1$, and $\varepsilon = 2.2$. The simulations were performed by means of the split-step fast-Fourier-transform method. The input optical field was taken as

$$u = A_0 \text{sech}\left(\frac{r-R}{w}\right)\exp(iM\theta), \tag{3}$$

where $A_0$, $w$, and $M$ represent, respectively, the amplitude, width, and integer topological charge (winding number) of the input, $(r,\theta)$ being the polar coordinates. Note that this input does not contain factor $r^{|M|}$, which usually appears in vortex *ansätze*. However, numerical simulations of the evolution initiated by input (3) demonstrate that the solution quickly restores it, similar to how it can be seen, e.g., in Figs. 7 and 8 of Ref. [25] for simulations of the standard cubic-quintic CGL equation (without the inhomogeneous diffusion term) with a similar Gaussian input and winding number $M = 1$ and 2.

To report results of the systematic simulations, we first vary the topological charge $M$ and relative amplitudes parameter of the effective diffusion $A_x$ (or $A_y$). Different ensuing propagation scenarios take place in domains plotted in Fig. 1(a) by fixing, first, equal strengths $A_x=A_y$ and equal oscillation periods $T_x=T_y$ of the inhomogeneous diffusion, for $\beta = 0.1$. The input produces instability [as shown in Fig. 1(b)] when strength $A_x$ falls below some critical values (underneath curve 1 in Fig. 1 (a)). In this case, the diffusion is too weak to support stable vortex propagation. If $A_x=$ $A_y$ take values between curves 1 and 2 in Fig. 1(a), the input evolves into a *gear-shaped* vortex soliton [Fig. 1(c)]. The number of lobes in the gear may be equal to $M$, as shown in Fig. 1(c), or it may be different from $M$, see Fig. 1(d).

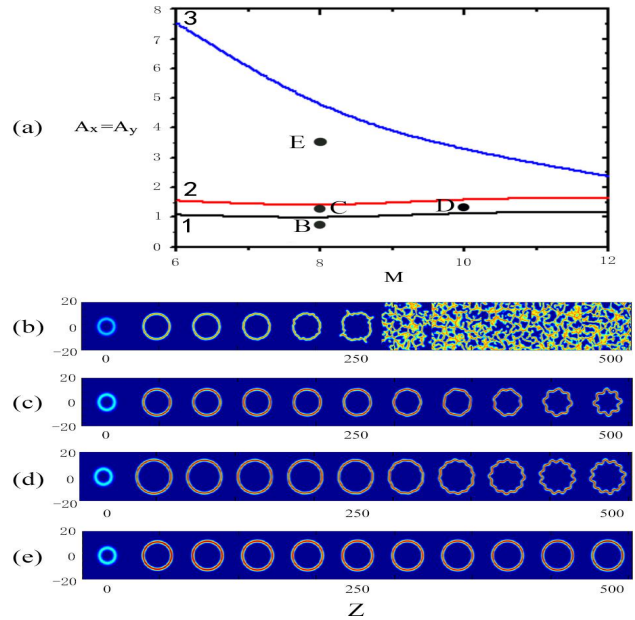

Fig. 1. (Color online) (a) Domains of different propagation scenarios of the input vortex field (3) for different values of $A_x = A_y$ and vorticity $M$, in the model with the symmetric diffusion modulation, $A_x = A_y$ and fixed $T_x = T_y = 1.0$, for the overall diffusion strength $\beta = 0.1$. Below curve 1: unstable (turbulent) propagation; between curves 1 and 2: generation of gear-shaped vortex solitons; between curve 2 and 3: generation of circular vortex soliton; above curve 3: decay of the wave field. (b) Unstable propagation for $M = 8$ and $A_x = A_y = 0.80$ [corresponding to point B in panel (a)]. (c) The generation of a gear-shaped vortex soliton for $M = 8$ and $A_x = A_y = 1.20$ [corresponding to point C in panel (a)]. (d) The generation of a gear-shaped vortex soliton for $M = 10$ and $A_x = A_y = 1.20$ [corresponding to point D in panel (a)]. (e) The creation of a circular vortex soliton for $M = 8$ and $A_x = A_y = 3.50$ [corresponding to point E in panel (a)].

Further, when $A_x = A_y$ take values between curves 2 and 3 in Fig. 1(a), the evolution produces a stable circular vortex soliton [see Fig. 1(e)], and when $A_x = A_y$ take values above curve 3 in Fig. 1(a), the input decays under the action of the strong diffusion effect (not shown here in detail).

Next we address the effect of the choice of unequal modulation periods $T_x$ and $T_y$ on the evolution of the input wave form, while still fixing equal relative strengths, $A_x = A_y$. To this end, we fix $\beta = 0.2$ and $M = 6$. When $T_y$ is too small [e.g., below curve 1 in Fig. 2(a), i.e., at $T_y < T_x$], the model generates into a quasi-square-shaped vortex soliton [at point B shown in panel (a)], as shown in Fig. 2(b). On the other hand, if $T_y$ is large enough to satisfy $T_y > T_x$, in the region between curves 1 and 2 in Fig. 2(a), it is shown in Fig. 2(c) that the input shrinks into an oval-shaped vortex soliton and tilts to the left [for instance, at point C shown in panel (a)].

For $T_y > T_x$ and $A_x = A_y$, the simulations always produce left-tilted oval-shaped vortices, as shown in Figs. 2 and 3 for $M = 6$ and $M = 8$, respectively. However, for $A_y > A_x$ and $T_y = T_x$, the oval vortices always appear to be right-tilted, as shown in Fig. 4. Under more general conditions, $A_y > A_x$ and $T_y > T_x$, the simulations produce both left-tilted and right-tilted oval-shaped vortices, as shown in Fig. 5, depending on the specific values of parameters $T_x$, $T_y$, $A_x$, and $A_y$. It is necessary to stress that the particular sign of the tilt is related to the choice of the positive vorticity in input (3), $M > 0$. Its opposite sign will generate the opposite sign of the tilt.

For larger values of $T_y$, above curve 2 in Fig. 2(a), the input loses its vorticity and transforms into fundamental soliton, as seen in Fig. 2(d) [which corresponds to point

D shown in panel (a)].

In Fig. 3, we further analyze the dynamics in the plane of ($T_y$,$T_x$) under the condition opposite to that presented in Fig. 2, namely, for $A_x=A_y$ and unequal values of $T_y$ and $T_x$, fixing $M=8$. Figure 3(b) shows the formation of a quasi-square-shaped vortex soliton. Figure 3(c) shows the generation of an elongated left-tilted oval vortex soliton. While these outcomes of the evolution are similar to those displayed in Fig. 2, a new result is displayed in Fig. 3(d), *viz.*, fission of the input into an obliquely oriented string composed of one fundamental and two vortex solitons. Comparing Fig. 2(a) and Fig. 3(a), we conclude that higher values of topological charge $M$ correspond to a larger region between curves 1 and 2, where the input evolves into a left-tilted oval-shaped vortex soliton.

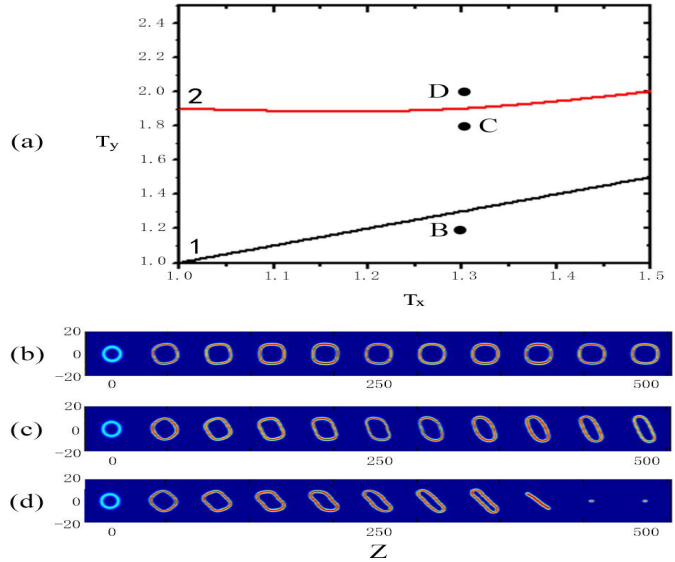

Fig. 2. (Color online) (a) Different propagation scenarios of input (3) with winding number $M = 6$ in parameter plane ($T_y$, $T_x$) under the conditions of $A_x = A_y = 2.0$ and $\beta = 0.2$. Below curve 1: the generation of a quasi-square-shaped vortex soliton; between curves 1 and 2: the formation of a left-tilted oval soliton; above curve 2: the transformation of the input into a single fundamental soliton. (b) The input evolves into a quasi-square-shaped vortex soliton for $T_x = 1.3$, $T_y = 1.2$ [corresponding to point B in panel (a)]. (c) The input shrinks into an ovalsoliton and tilts to the left, for $T_x = 1.3$, $T_y = 1.8$ [corresponding to point C in panel (a)]. (d) The input decays into a single fundamental soliton for $T_x = 1.3$, $T_y = 2.0$ [corresponding to point D in panel (a)].

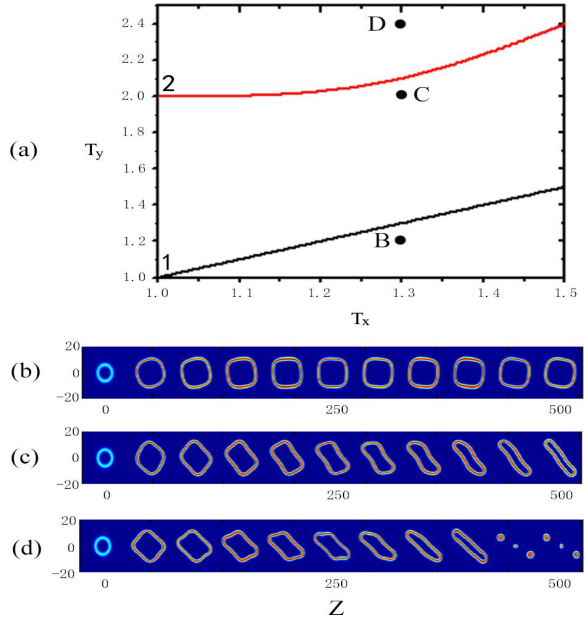

Fig. 3. (Color online) (a) Different propagation scenarios for input (3) with $M = 8$, in the parameter plane $(T_y, T_x)$, the other parameters being the same as in Fig. 2. Below curve 1: The generation of a quasi-square-shaped vortex soliton; between curves 1 and 2: The formation of a left-tilted oval vortex soliton; above curve 2: The transformation into a set of one fundamental and two vortex solitons. (b) The input evolves into a quasi-square-shaped soliton at $T_x = 1.3$ and $T_y = 1.2$ [corresponding to point B in panel (a)]. (c) The soliton shrinks into an elongated left-tilted oval vortex soliton at $T_x = 1.3$ and $T_y = 2.0$ [corresponding to point C shown in panel (a)]. (d) The input splits into a fundamental soliton and two vortex satellites, which form an oblique string, at $T_x = 1.3$ and $T_y = 2.4$ [corresponding to point D in panel (a)].

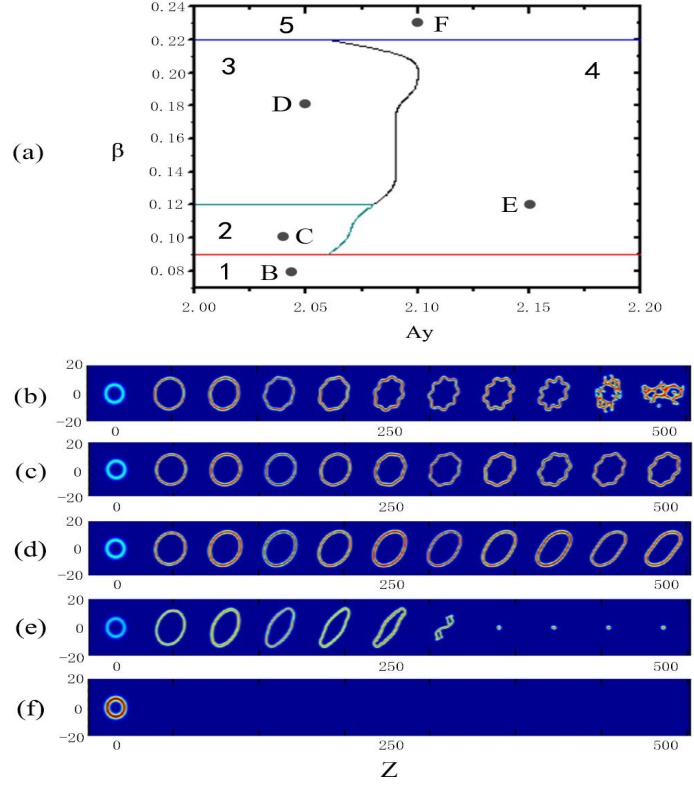

Fig. 4. (Color online) (a) Different propagation scenarios for vortex input with $M = 8$ in the parameter plane ($\beta$, $A_y$), under conditions $A_y > A_x = 2.0$ and $T_y = T_x = 1.4$. Region 1: the input develops instability; Region 2: the input evolves into a gear-shaped right-tilted oval vortex soliton; Region 3: the input evolves into a right-tilted oval vortex soliton; Region 4: the input evolves into a single fundamental soliton; Region 5: the input decays. (b) The instability which eventually sets in for $A_y = 2.04$ and $\beta = 0.08$ [corresponding to point B in panel (a)]. (c) The generation of the right-tilted gear-shaped oval vortex for $A_y=2.03$ and $\beta = 0.10$ [corresponding to point C in panel (a)]. (d) The generation of the right-tilted oval vortex soliton for $A_y=2.05$ and $\beta = 0.18$ [corresponding to point D in panel (a)]. (e) The input evolves into a single fundamental soliton for $A_y = 2.15$ and $\beta = 0.12$ [corresponding to point E in panel (a)]. (f) Decay of the field for $A_y = 2.10$ and $\beta = 0.23$ [at point F in panel (a)].

Next, in Fig. 4 we display dynamical scenarios in the parameter plane ($\beta$, $A_y$) under condition $A_y > A_x$ and for $T_y = T_x = 1.4$. In region 1, as shown in Fig. 4(a), the input originally evolves into a right-handed gear-shaped vortex, which is eventually subject to strong instability, see Fig. 4(b). When the overall diffusion strength, $\beta$, increases, taking values in region 2 in Fig. 4(a), the input evolves into a quasi-stable right-handed gear-shaped vortex soliton, as shown in Fig. 4(c), corresponding to point C in Fig. 4(a). Further, when $\beta$ takes still larger values in region 3 in Fig. 4(a), the input evolves into a quasi-stable oval vortex soliton, as displayed in Fig. 4(d), corresponding to point D in Fig. 4(a).

The effect of $\beta$ on the propagation dynamics is clearly illustrated by Fig. 4: a larger value of $\beta$ produces a higher friction force that helps to maintain robust propagation of the solitary modes (see panels (c) and (d) in Fig. 4) for $z = 500$. At very large values of $z$, both the gear-shaped and plain-oval-shaped right-tilted vortex solitons, displayed in Figs. 4(c) and (d), respectively, gradually increase the tilt and develop instability. Eventually, the former structure evolves into an irregular state, while the latter one transforms into a single fundamental soliton.

Furthermore, if $A_y$ is large enough to get into region 4 shown in Fig. 4(a), the input evolves into a fundamental soliton, as displayed in Fig. 4(e), which corresponds to point E in Fig. 4(a). Then, if $\beta$ is large enough, that is, in region 5 shown in Fig. 4(a), the input completely decays, as illustrated in Fig. 4(f), which corresponds to point F in Fig. 4(a).

Finally, we consider the more general case, with $A_y > A_x$ and $T_y > T_x$. The domains corresponding to different propagation scenarios in the plane of $(A_y, T_y)$ for $\beta = 0.20$ are shown in Fig. 5(a). For small values of $A_y$, the input is distorted and into a left-tilted oval one, as shown in Fig. 5(b). At larger values of $A_y$ between curves 1 and 2 in Fig. 5(a), the input shrinks into a regular right-tilted oval vortex soliton, as shown in Fig. 5(c). Thus, the evolving shape gradually tilts from left to right, developing the oval shape, with the increase of $A_y$. Eventually, the tilt attains a limit value, ceasing to change with the further growth of $A_y$. If $A_y$ is sufficiently large to get into the domain above curve 2 in Fig. 5(a), the oval-shaped vortex soliton gradually shrinks, keeping the constant tilt, and at large $z$ it splits into a string of three fundamental solitons, which are aligned in the same direction, see Fig. 5(d).

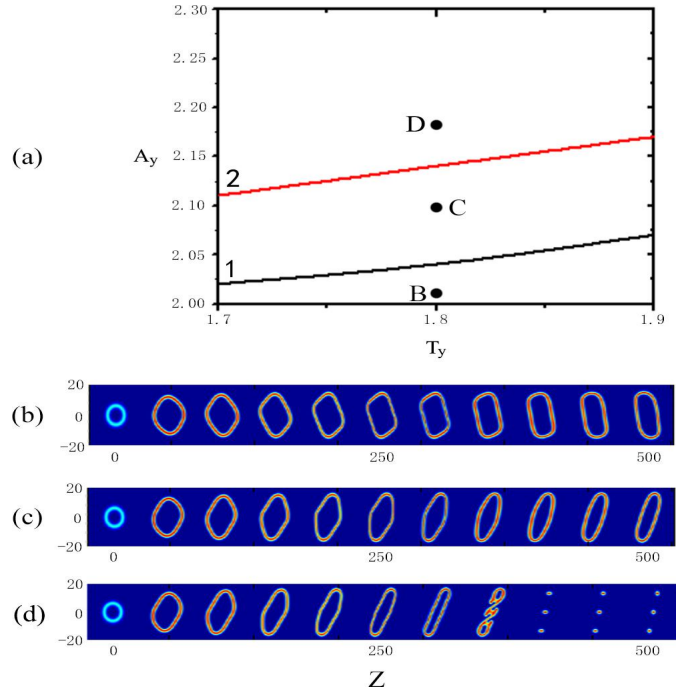

Fig. 5. (Color online) (a) Different propagation scenarios for vortex input (3) with $M = 8$, in the plane of $(A_y, T_y)$ in the case of $A_y > A_x = 2.0$ and $T_y > T_x = 1.0$ for $\beta = 0.20$. (b) The input evolves into a left-tilted quasi-oval vortex soliton at $T_y=1.8$ and $A_y=2.01$ [corresponding to point B in panel

(a) ]; (c) The input evolves into an oval right-leaning vortex for $T_y$=1.8 and $A_y$=2.10 [corresponding to point C in panel (a) ]; (d) An example of the fission of input into an oblique string composed of three fundamental solitons for $T_y$=1.8 and $A_y$=2.18 [corresponding to point D in panel (a) ].

## 4 Conclusions

The objective of this work is to demonstrate that the use of the effective diffusion (spatial filtering, in terms of optics), subject to spatially periodic modulation, makes it possible to create new species of stable ring-shaped vortex solitons in dissipative media modeled by the two-dimensional CGL (complex Ginzburg-Landau) equation with the cubic-quintic nonlinearity. These possibilities are further expanded by the use of anisotropic diffusion. By means of numerical methods, we demonstrate the generation of gear- and square-shaped vortex rings, as well as prolate obliquely oriented oval rings, from a localized two-dimensional input with internal vorticity. In addition, string-shaped patterns built of a central fundamental soliton and two vortex satellites, or of three fundamental solitons, can be generated too, depending on parameters of the CGL equation and on input state (including its vorticity). Stability domains for these types of the outputs are also given in the model's parameter space.

**Acknowledgments** This work was supported by the National Natural Science Foundations of China (Grant Nos. 11174061, 61675001, and 11774068), the Guangdong Province Nature Foundation of China (Grant No. 2017A030311025), and the Guangdong Province Education Department Foundation of China (Grant No. 2014KZDXM059). We declare that we do not have any conflict of interest in connection with the present work.